\documentclass[sigconf]{acmart}

\usepackage{graphicx}
\usepackage{latexsym}
\usepackage[utf8]{inputenc} 
\usepackage[T1]{fontenc}    
\usepackage{url}            
\usepackage{booktabs}       
\usepackage{nicefrac}       
\usepackage{microtype}      
\usepackage{latexsym}
\usepackage{xspace}
\usepackage{dsfont}
\usepackage{multirow}
\usepackage{subcaption}
\usepackage{amsmath}
\usepackage[prependcaption]{todonotes}
\usepackage{wrapfig,lipsum}
\usepackage{csquotes}
\usepackage{xcolor}
\usepackage{fancyvrb}
\usepackage[export]{adjustbox}
\usepackage{arydshln}
\usepackage{colortbl}
\usepackage{array}

\usepackage{navid}
\usepackage{bm}

\def\eqref#1{equation~\ref{#1}}

\def\1{\bm{1}}

\DeclareMathAlphabet{\mathsfit}{\encodingdefault}{\sfdefault}{m}{sl}
\SetMathAlphabet{\mathsfit}{bold}{\encodingdefault}{\sfdefault}{bx}{n}

\newcommand{\modellm}{QL\xspace}
\newcommand{\modelmp}{MatchPyramid\xspace}
\newcommand{\modelknrm}{KNRM\xspace}
\newcommand{\modelcknrm}{ConvKNRM\xspace}
\newcommand{\modeltk}{TK\xspace}
\newcommand{\modelbert}{BERT\xspace}
\newcommand{\modelss}{Seq2SeqAttention\xspace}
\newcommand{\modelpgn}{PGN\xspace}
\newcommand{\modeltpgn}{T-PGN\xspace}
\newcommand{\modeltt}{Transf2Transf\xspace}
\newcommand{\modelbertt}{BERT2Transf\xspace}

\newcommand{\lossce}{\texttt{CE}\xspace}
\newcommand{\losslul}{\texttt{NL3U}\xspace}
\newcommand{\lossnegl}{\texttt{NLL}\xspace}
\newcommand{\lossmxmrg}{\texttt{Marg}\xspace}

\copyrightyear{2021} 
\acmYear{2021} 
\setcopyright{acmlicensed}\acmConference[ICTIR '21]{Proceedings of the 2021 ACM SIGIR International Conference on the Theory of Information Retrieval}{July 11, 2021}{Virtual Event, Canada}
\acmBooktitle{Proceedings of the 2021 ACM SIGIR International Conference on the Theory of Information Retrieval (ICTIR '21), July 11, 2021, Virtual Event, Canada}
\acmPrice{15.00}
\acmDOI{10.1145/3471158.3472229}
\acmISBN{978-1-4503-8611-1/21/07}

\definecolor{mint}{rgb}{0.24, 0.71, 0.54}

\begin{document}
\fancyhead{}
\fancyfoot{}
\title{A Modern Perspective on Query Likelihood with Deep Generative Retrieval Models}


\author{Oleg Lesota}
\email{oleg.lesota@jku.at}
\affiliation{%
  \institution{Johannes Kepler University Linz}
  \institution{Linz Institute of Technology, AI Lab}
  \city{Linz}
  \country{Austria}
  }

\author{Navid Rekabsaz}
\email{navid.rekabsaz@jku.at}
\affiliation{%
  \institution{Johannes Kepler University Linz}
  \institution{Linz Institute of Technology, AI Lab}
  \city{Linz}
  \country{Austria}
  }

\author{Daniel Cohen}
\email{daniel_cohen@brown.edu}
\affiliation{%
  \institution{Brown University}
  \city{Providence, R.I.}
   \country{USA}
   }

\author{Klaus Antonius Grasserbauer}
\email{klaus.grasserbauer@jku.at}
\affiliation{%
  \institution{Johannes Kepler University Linz}
  \city{Linz}
  \country{Austria}
  }

\author{Carsten Eickhoff}
\email{carsten@brown.edu}
\affiliation{%
  \institution{Brown University}
  \city{Providence, R.I.}
   \country{USA}
   }

\author{Markus Schedl}
\email{markus.schedl@jku.at}
\affiliation{%
  \institution{Johannes Kepler University Linz}
  \institution{Linz Institute of Technology, AI Lab}
  \city{Linz}
  \country{Austria}
  }


\begin{abstract}
\vspace{-0.1cm}

Existing neural ranking models follow the text matching paradigm, where document-to-query relevance is estimated through predicting the matching score. Drawing from the rich literature of classical generative retrieval models, we introduce and formalize the paradigm of \textit{deep generative retrieval models} defined via the cumulative probabilities of generating query terms. This paradigm offers a grounded probabilistic view on relevance estimation while still enabling the use of modern neural architectures. In contrast to the matching paradigm, the probabilistic nature of generative rankers readily offers a fine-grained \textit{measure of uncertainty}. We adopt several current neural generative models in our framework and introduce a novel generative ranker (\textit{T-PGN}), which combines the encoding capacity of Transformers with the Pointer Generator Network model. We conduct an extensive set of evaluation experiments on passage retrieval, leveraging the MS~MARCO Passage Re-ranking and TREC Deep Learning 2019 Passage Re-ranking collections. Our results show the significantly higher performance of the T-PGN model when compared with other generative models.
Lastly, we demonstrate that exploiting the uncertainty information of deep generative rankers opens new perspectives to query/collection understanding, and significantly improves the \textit{cut-off prediction} task. 

\vspace{-0.2cm}
\end{abstract}



\keywords{\vspace{-0.1cm}Neural IR;
generative ranking model;
uncertainty;
cut-off prediction}


\maketitle

\section{Introduction}
\label{sec:introduction}
Neural ranking models have yielded remarkable improvements to information retrieval (IR) by estimating a highly non-linear function of relevance of a query to a document. 
Arguably, all existing neural ranking models~\cite{Hofstaetter2020_sigir,nogueira2019document,dai2018convolutional,fan2018modeling,xiong2017end,mitra2017learning,guo2016deep,lu2013deep} follow the text matching paradigm, where relevance is calculated as the predicted matching score of each document to a given query. In this sense, these neural ranking models appear to be the descendants of matching-based (or similarity-based) models such as the vector space model~\cite{salton1975vector}, where the model estimates the relevance score of a document $D$ to a query $Q$ by the matching function $f(Q,D)$. We refer to these models (whether neural or classical ones) as \emph{matching models}. 

A generative view on IR was first introduced by \citet{ponte1998language}, where -- unlike in matching models -- relevance is expressed in terms of a conditional probability in a well-formed probabilistic framework. In particular, the query likelihood language model estimates relevance as the probability of the query being generated by a language model of the document, namely $P(Q|\Phi_D)$. This regime provides a powerful probabilistic framework to IR, and has been the base for numerous approaches (see \citet{zhai2008statistical} for further details). Our paper provides a modern perspective on the fundamental principle of the generative paradigm for IR through the recent advancements in deep generative models. We introduce and provide the theoretical foundations of \emph{deep generative ranking models}, comprehensively study the characteristics and performance of the models' various architectural choices for passage retrieval, and show the immediate benefits of the probabilistic nature of this paradigm in providing more-than-relevance information.

\begin{figure*}[t]
\centering
\subcaptionbox{Representation-based (e.g., DSSM)\label{fig:repr}}{\includegraphics[width=0.18\textwidth]{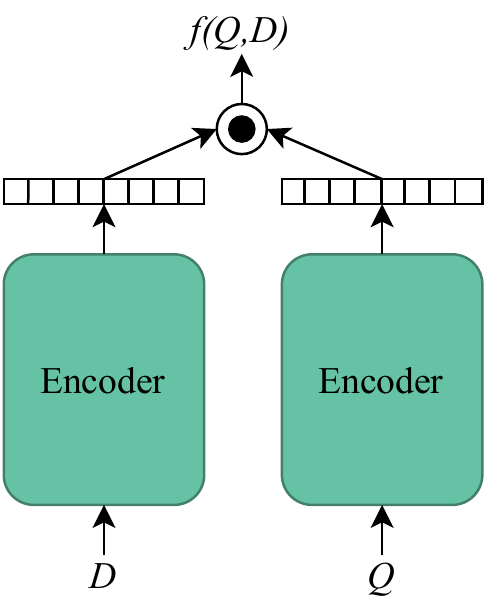}}
\hspace{8mm}
\subcaptionbox{Query-Document Interaction (e.g., KNRM, BERT)\label{fig:inter}}{\includegraphics[width=0.19\textwidth]{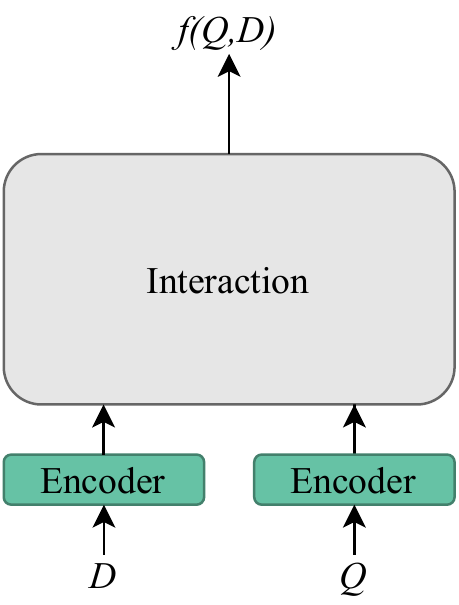}}
\hspace{8mm}
\subcaptionbox{Deep Generative Models\label{fig:gen}}{\includegraphics[width=0.19\textwidth]{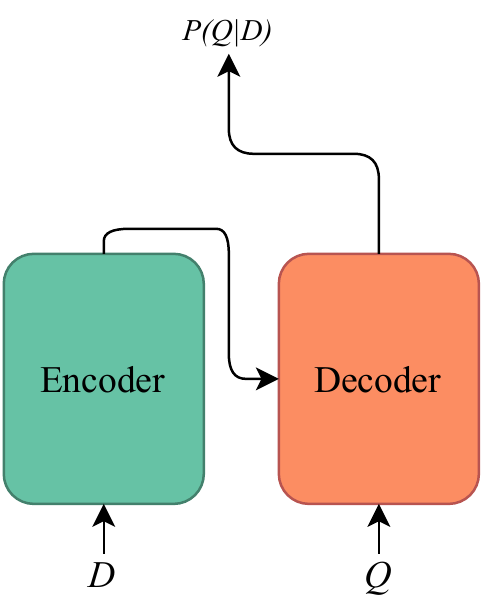}}
\centering
\caption{Schematic views of representation-based and query-document interaction models with representative examples of models (the two categories of the matching paradigm), and deep generative ranking models (the generative paradigm).}
\label{fig:paradigms}
\vspace{-0.2cm}
\end{figure*}

Let us first discuss the overall architectural differences/similarities across various IR models, as highlighted in Figure~\ref{fig:paradigms}. \textit{Representation-based models} first encode query and document into separate embeddings, and then use these embeddings to calculate query-to-document relevance scores~\cite{hu2014convolutional,shen2014learning,huang2013learning}. In \textit{query-document interaction models}, the query-term-to-document-term interactions (in the form of similarities or attention networks) are used to create a final feature vector, and hence estimate the relevance score~\cite{Hofstaetter2020_sigir,dai2018convolutional,fan2018modeling,xiong2017end,mitra2017learning,hui2017pacrr,pang2016text,guo2016deep,hu2014convolutional,lu2013deep}. In this category of models, large-scale pre-trained language models such as BERT~\cite{devlin2019bert} have shown significant improvements in retrieval performance~\cite{nogueira2019passage,macavaney2019contextualized}. 

Deep generative ranking models (Figure~\ref{fig:gen}) view relevance estimation as the probability of generating the query, given the document. Models follow the sequence-to-sequence (seq2seq) encoder-decoder architecture~\cite{sutskever2014sequence}. The model first encodes the document, and then uses the encoded embeddings to provide probability distributions over the space of possible queries at the output of the decoder. This framework in addition to effective estimation of relevance, provides a distinctive benefit in comparison with other retrieval models: the probabilistic nature of generative models enables the extraction of actionable information in addition to mere relevance scores. This probabilistic information for example can take the form of uncertainty estimates of the model. Such uncertainty estimation is directly achieved from the output of the model and does not need any model modification, nor does it impose any extra computation. As we show in the present work, this uncertainty information can effectively be exploited for better understanding of the underlying collection and training data, and also in downstream tasks such as rank cut-off prediction. 

In addition, the decoupling of query/document  encoding in the architecture of deep generative model enables the ranking model to store document embeddings at index time and later exploit them at inference time (similar to representation-based models). At the same time, the decoder embeddings still effectively interact with encoded embeddings (typically through attention mechanisms), which is analogous to interaction-focused models. The use of attention mechanisms over the document during decoding facilitates effective interaction with encoded document embeddings (as in query-document interaction models), but also enables the potential incorporation of orthogonal notions, such as personalization, diversity or fairness into the model.


The present work explores various aspects of the generative IR paradigm from the perspective of deep generative models. Concretely, we first formalize the theoretical connection between the introduced deep generative ranking models and classical generative IR models, in particular the query likelihood language model. 

We then investigate the effectiveness of various generative architectures in passage retrieval. To this end, we conduct a comprehensive study on the use of the state-of-the-art neural language generation techniques for retrieval. 
We study various models, among them Pointer Generator Networks (\modelpgn)~\cite{see2017get}, and a recently proposed combination of BERT and Transformers~\cite{liu2019text}. In addition to these models, we combine the benefits of \modeltpgn in query decoding with those of BERT in document encoding, and propose a new generative ranking model referred to as \modeltpgn. We evaluate these generative models on the MS~MARCO Passage Re-ranking~\cite{nguyen2016ms} and the TREC Deep Learning 2019 Passage Re-ranking task~\cite{craswell2020overview}. The results demonstrate that among the generative models, our introduced \modeltpgn model shows the overall best performance.



Finally, drawing from the probabilistic framework of deep generative models, we calculate a measure of uncertainty reflecting the model's confidence in the prediction of a given query. The uncertainty estimate is achieved by calculating the entropy values of the probability distributions of query term generation. We use the resulting uncertainty estimates to first analyze the existence of bias with respect to term positions in the queries of MS~MARCO and then exploit this extra information for cut-off prediction, observing a significant improvement in the task.

\noindent To summarize, our main contribution is four-fold:
\vspace{-0.1cm}
\begin{itemize}
    \item Introducing the novel deep generative ranking models and formalizing them in the perspective of classical generative IR models.
    \item Adopting several recent deep generative language models for ranking, and introducing a new generative ranker (T-PGN).
    \item Conducting a large set of evaluation experiments on various generative models for passage retrieval.
    \item Showcasing the potential of deep generative ranking models for uncertainty estimation of relevance scores and its use in a cut-off prediction task.
\end{itemize}
\vspace{-0.1cm}


The paper is organized as follows: Section~\ref{sec:related} reviews related literature. In Section~\ref{sec:method}, we introduce the deep generative ranking models and explain various architectural choices as well as their potential for uncertainty estimation. Section~\ref{sec:setup} describes our design of experiments, whose results are reported and discussed in Section~\ref{sec:results}. The accompanying source code is available at \url{https://github.com/CPJKU/DeepGenIR}.


\vspace{-0.2cm}
\section{Related Work}
\label{sec:related}

\subsection{Neural Retrieval Models}
In the category of query-document interaction models, we can distinguish between three groups of models. The first group captures patterns of similarity values across terms that appear close together within the query and within the document~\cite{fan2018modeling,hui2018co,hui2017pacrr,pang2016text}. The second group captures patterns of frequencies across ranges of similarity values~\cite{Hofstaetter2020_sigir,dai2018convolutional,hofstatter2019effect,xiong2017end,guo2016deep}. The last ones are based on large-scale pre-trained language models, as the use of these models has shown significant performance gains in various IR tasks. 

For instance, the BERT model is used for document/passage retrieval through fine-tuning~\cite{nogueira2019passage}, combining them with other ranking models~\cite{macavaney2019contextualized}, expanding to other more efficient variations~\cite{macavaney2020efficient} or dense retrieval approaches~\cite{xiong2021approximate,khattab2020colbert}. In this paper, we also investigate the benefits of exploiting such large-scale pre-trained language models in the context of deep generative ranking models.

Finally, in addition to the mentioned neural models, other studies exploit the inherent efficiency of  classic IR models while aiming to improve their effectiveness using pre-trained embedding models. This is done for instance by generalizing term salience with translation models~\cite{rekabsaz2016generalizing,rekabsaz2017exploration}, and re-weighting terms~\cite{zheng2015learning}, or through adapting word embeddings for document retrieval by post-filtering~\cite{rekabsaz2017word}, retrofitting~\cite{hofstaetter2019enriching}, or re-training on local documents~\cite{diaz2016query}. 

\vspace{-0.2cm}
\subsection{Neural Generative Models in IR}

Neural generative models have been utilized in various IR tasks. As examples, \citet{zamani2020generating} study the use of a seq2seq model to generate queries, whose results are used as a source of weak supervision for asking clarifying questions. \citet{ren2018conversational} and later \citet{yang2019hybrid} approach the task of reformulating conversational queries into search-engine-friendly queries using seq2seq with attention models. \citet{ahmad2019context,ahmad2018multi} use a similar generative model to train a query recommender, which facilitates the reformulation of users' queries and hence the effective ranking of documents. 

In the context of neural ranking models, \citet{nogueira2019document} use a seq2seq Transformer model~\cite{vaswani2017attention,zerveas2019brown} to expand documents with generated queries, and adopt a BERT-based matching model to conduct retrieval on the expanded documents. In a more recent work, \citet{nogueira2020document} exploit the T5 model~\cite{raffel2020exploring} (a pre-trained Transformer-based seq2seq model) to perform binary classification of query-document pairs as relevant or irrelevant. In this approach, query and document are both given as the input to the encoder, and the generated output of the decoder is two possible tokens (``true'' or ``false'') corresponding to the relevant and non-relevant class. The authors use the logits corresponding to the two tokens to infer the relevance score. Based on the discussion in Section~\ref{sec:introduction} and on Figure~\ref{fig:paradigms}, despite the seq2seq architecture of this model, this approach can in fact be categorized among the query-document interaction models, since the input is the concatenation of query and document, and the output is their relevance score. In contrast, the deep generative models presented in the work at hand generate text queries from input documents, where the probability of generating each query term is defined over all possible words (and not over two tokens). Parallel to our work, \citet{QLZhuangL2021} investigate query likelihood models built upon a Transformer-based seq2seq architecture. Our work expands their study by investigating a wide range of deep generative architectures in IR ranking, and showing the fundamental benefits of generative ranking models for query understanding and in downstream tasks.



In a larger context, neural language generation models span various language processing tasks, i.e.\ machine translation~\cite{vaswani2017attention}, abstractive document summarization~\cite{liu2019text,see2017get}, dialogue generation~\cite{vinyals15a}, and question answering~\cite{dos2020beyond,xing2015normalized}. The present work benefits from and contributes to these studies by adopting deep generative models in the context of retrieval, and introducing a new generative model.

\vspace{-0.2cm}
\section{Deep Generative Ranking Models}
\label{sec:method}
In the following, we first formulate deep generative ranking models by highlighting their connections to classical generative models~\cite{ponte1998language,zhai2008statistical,Ai2019NeuralGM}. We then describe the various loss functions used to train the models, followed by a detailed description of the proposed T-PGN and other generative ranking models used in this study. Finally, we introduce our approach to uncertainty estimation defined on the probability space resulting from deep generative rankers.

\vspace{-0.2cm}
\subsection{Definition}
\citet{ponte1998language} introduced the language modeling approach to IR and proposed a new scoring model based on this approach, which later has been called the query likelihood model (\modellm). The language modeling approach defines the relevance of document $D$ to query $Q$ based on the conditional probability $P(Q|D)$.\footnote{More precisely, $P(Q|D,R=r)$, i.e.~the probability of $Q$ given $D$ and a level of relevance $r$.} This probability is rooted in the idea that a user who wants to find document $D$ would utilize query $Q$ to retrieve the document~\cite{zhai2008statistical}. 
$P(Q|D)$ is defined by $P(Q|\Phi_D)$ -- the probability of generating query $Q$ using the language model $\Phi_D$, built based on document $D$. \citet{zhai2008statistical} explains the objective of $\Phi_D$ as ``modeling the queries that a user would use in order to retrieve documents'', highlighting the fact that, although $\Phi_D$ is a document language model, it is effectively a model meant for queries and not for documents.

The most well-known way to use the language modeling approach is by utilizing a multinomial language model assuming query term independence~\cite{zhai2008statistical}, resulting in the following model formulation, known as QL:
\begin{equation}
\text{\modellm:} \quad P(Q|D) \approx P(Q|\Phi_D) = \prod_{q_i \in Q} P(q_i|\Phi_D)
\label{eq:qlm}
\end{equation}
The language model $\Phi_D$ is commonly defined as a unigram probability distribution of $D$ over all terms in the vocabulary, smoothed using the collection as background statistics. The relevance score of query to document is defined as the logarithm of the conditional probability.
\begin{equation}
\text{\modellm:} \quad \text{score}(Q, D) = \sum_{q_i \in Q} \log P(q_i|\Phi_D)
\label{eq:qlm:score}
\end{equation}

Deep generative ranking models follow a similar perspective to relevance estimation as the \modellm: Document $D$ should be scored as more relevant to query $Q$ if the model assigns a higher value to the probability of generating $Q$ when conditioned on $D$. This probability is calculated based on an encoder-decoder architecture. The encoder receives document $D$ as a sequence of input tokens and provides a (contextualized) representation of the document. The decoder uses the document's representation as well as previous query tokens and estimates as output the probability of generating the next query token. The decoder is in fact a \emph{query language model} which outputs the probability of generating the next query token, conditioned on the representation of the document in auto-regressive fashion (one after another). Given such a generative model, the relevance score is defined as the probability of generation of the query, conditioned on the document. The generation probability is formulated as follows:

\begin{equation}
P_\theta(Q|D) = \prod_{q_i \in Q} P_\theta(q_i|D,q_{j<i})
\label{eq:nglm}
\end{equation}
where $q_{j<i}$ denotes the query tokens preceding the current token, and $\theta$ indicates the model's parameters learned using training data. Similar to \modellm, the relevance score is defined as the logarithm of the conditional probability:
\begin{equation} 
\text{score}(Q,D) = \sum_{q_i \in Q} \log P_\theta(q_i|D,q_{j<i})
\label{eq:nglm:score}
\end{equation}

Having outlined the conceptual similarities of deep generative ranking models and \modellm, let us now discuss the differences, particularly by comparing the formulations in Eq.~\ref{eq:qlm} and Eq.~\ref{eq:nglm}. One difference is that deep generative models are not constrained by the term independence assumption, as the generation of each token is conditioned on the previous terms. Another difference is rooted in the language models that deep generative models use to generate queries. While \modellm utilizes $\Phi_D$ -- the language model of document $D$ -- deep generative models use the query language model that is created by observing all queries in the training data. In fact, in contrast to \modellm, deep generative ranking models explicitly train a language model for queries, whose generation probabilities are conditioned on the given document. In this sense, deep generative ranking models can be seen as an alternative implementation of the language modeling approach to IR, while still benefiting from the advantages of neural ranking models.

\vspace{-0.2cm}
\subsection{Ranking Loss Functions}
Considering the provided formulation of deep generative ranking models, we discuss in this section the training loss functions used to train generative models. Given a pairwise learning-to-rank setting, the training data of retrieval models is provided in the form of a query $Q$ with a relevant and a non-relevant document, denoted as $D^{+}$ and $D^{-}$ respectively.

The first loss is the well-known \emph{negative log likelihood} (\lossnegl), commonly used to train generative models, and defined as follow:
\begin{equation}
\mathcal{L}_{\lossnegl} = - \sum_{(Q,D^{+}) \in \mathcal{T}}{\log P_\theta(Q|D^{+})}
\label{eq:nll}
\end{equation}
where $\mathcal{T}$ denotes the collection of training data. It is evident that \lossnegl only considers the relevant document and does not use the non-relevant one. The next loss function is \emph{margin ranking} (\lossmxmrg) formulated below:
\begin{equation}
\mathcal{L}_{\lossmxmrg} = \sum_{(Q,D^{+},D^{-}) \in \mathcal{T}}{\max\{0, b - \log P_\theta(Q|D^{+}) + \log P_\theta(Q|D^{-})\}} 
\label{eq:mxmrg}
\end{equation}
The \lossmxmrg loss increase the differences between the predicted relevance score of the relevant document and the predicted relevance score of the non-relevant document up to a margin threshold $b$. This loss can accept relevance scores in any range and is therefore commonly adopted in ranking models.



Our final loss function proposed by \citet{dos2020beyond} expands \lossnegl by adding the unlikelihood probability of negative documents, namely the logarithm of one minus probability of the negative document in training data. We refer to this loss as \emph{negative log likelihood log unlikelihood} (\losslul), defined as follows:
\begin{equation}
\mathcal{L}_{\losslul} = - \sum_{(Q,D^{+}) \in \mathcal{T}}{\log P_\theta(Q|D^{+}) + \log (1 - P_\theta(Q|D^{-}))}
\label{eq:lul}
\end{equation}

\subsection{Neural Generative Ranking Architectures}
Based on our formulation of neural generative ranking models, any neural generative model can be exploited for retrieval, namely by calculating the query-to-document relevance estimated from the generation probability distributions. In the following, we first briefly describe the generative models studied in this paper, and then explain our proposed \modeltpgn model. These models are selected based on their strong performance in tasks such as abstractive document summarization and machine translation. 

\begin{figure}[t]
\centering
\includegraphics[width=0.5\textwidth]{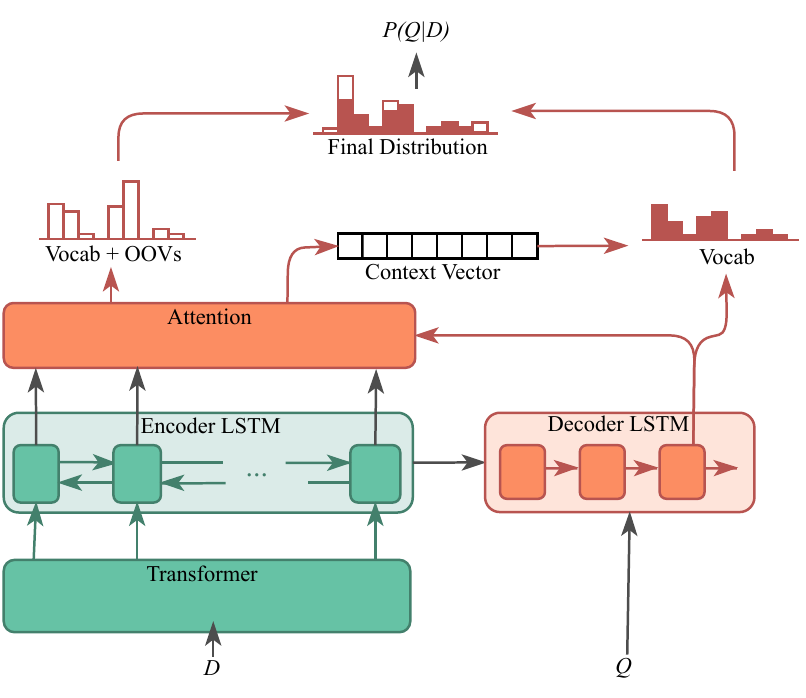}
\centering
\vspace{-0.3cm}
\caption{Transformer Pointer Generator Network (\modeltpgn)}
\label{fig:tpgn}
\vspace{-0.5cm}
\end{figure}

\vspace{-0.1cm}
\paragraph{\modelss} The Sequence-to-Sequence with Attention model~\cite{nallapati2016abstractive} is an extension to the baseline Sequence-to-Sequence model~\cite{sutskever2014sequence}. The baseline model consists of an encoder LSTM and a decoder LSTM, where the last hidden state of the encoder is given to the initial hidden state of the decoder. \modelss extends this models by the attention network, defined on the encoder hidden states and conditioned on the hidden state of the decoder LSTM. This attention mechanism enables the immediate access of the decoder to all document embeddings at encoder, facilitating information flow in the model. 

\vspace{-0.1cm}
\paragraph{\modelpgn} \citet{see2017get} introduce the Pointer Generator Network, which expands the \modelss model by a novel copy mechanism. The objective of this copy mechanism is to facilitate the transfer of the out-of-vocabulary (OOV) terms appearing in the document directly to the output query. This approach has shown highly competitive performance in abstractive summarization benchmarks. This is due to the fact that in summarization (similar to IR) rare words -- which are commonly removed from the list of vocabularies due to their low collection frequencies -- can be highly salient, and hence crucial for the success of the task.

\vspace{-0.1cm}
\paragraph{\modeltt} The Transformer-to-Transformer is introduced by \citet{vaswani2017attention} in the context of machine translation. The model consists of multiple layers of encoder Transformers to contextualize document embeddings with self-attention, followed by multiple layers of decoder Transformers. The decoder Transformers generate output probability distributions by contextualizing query embeddings and attending to the final embeddings of the encoder.

\vspace{-0.1cm}
\paragraph{\modelbertt} The BERT-to-Transformer model, recently introduced by \citet{liu2019text}, achieves state-of-the-art results on abstractive text summarization collections. The model has a similar architecture to the one of \modeltt but instead of Transformers uses a BERT model encoder.  

\vspace{-0.1cm}
\paragraph{\textbf{Transformer Pointer Generator Networks (\modeltpgn)}} We introduce the Transformer Pointer Generator Networks model (\modeltpgn) which combines the advantages of the \modelpgn model with Transformers. The architecture of the model is shown in Figure~\ref{fig:tpgn}. The \modeltpgn model provides a multi-layer encoder Transformer to create contextualized word embeddings of document terms. These embeddings are then passed to the encoder LSTM, whose final hidden state is used as the initial state of the decoder LSTM. Similar to \modelpgn, the attention distribution over the contextualized document embeddings (containing the OOV terms) is combined with the output distribution provided by the decoder to form the final distribution. This provides a probability distribution of query generation defined over all words in the vocabulary as well as the OOV terms appearing in document. 

\vspace{-0.2cm}
\subsection{Uncertainty Estimation in Neural Generative Rankers}\label{sec:method_uncertainty}
\begin{figure}[t]
\centering
\includegraphics[width=0.30\textwidth]{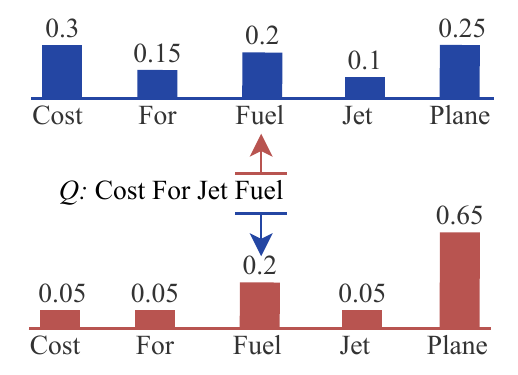}
\centering
\vspace{-0.2cm}
\caption{Illustration of different uncertainty estimates on a given query term position. While the probability of generating the term \emph{Fuel} is the same in both cases, the upper distribution contains a much higher degree of uncertainty.}
\label{fig:uncertainty_ex}
\vspace{-0.5cm}
\end{figure}

Given a document, deep generative models predict a probability distribution for each term of a query. As discussed in Section~\ref{sec:introduction}, this probabilistic perspective enables the calculation of the uncertainty of the model with respect to this prediction. In the following, we explain our approach to calculating an uncertainty estimate given any deep generative model.  



At every step $i$ of query term generation, the deep generative models estimate a probability distribution over all terms of the vocabulary. Despite the selected probability value for the term in the position $i$, $P_\theta(q_i|D,q_{j<i})$, the form of the predicted probability distribution reveals parallel information about the model. In fact, the same generation probability of a term may result from different kinds of probability distributions. This point is illustrated with a toy example in Figure~\ref{fig:uncertainty_ex} for the term \emph{Fuel}. As shown, if the distribution of the term generation probabilities is close to uniform (the upper graphic in Figure~\ref{fig:uncertainty_ex}), the model is not certain about the generation probability, as many terms have comparable chance to be generated in the next position. In contrast, when the distribution is more skewed, the model is more certain about possible generation terms (the lower graphic in Figure~\ref{fig:uncertainty_ex}). Despite these different distributions, the predicted probability values of \emph{Fuel} in both distributions are equal. In fact, this \textit{term-level uncertainty} provides extra information that might not be captured in the predicted probability values, and hence the predicted relevance score. 

Similar to \citet{xu2020understanding}, we define term-level uncertainty as the entropy of the \emph{nucleus} probability distribution at each step. The nucleus distribution~\cite{holtzman2020curious} provides a well-behaved version of the original generation probability distribution, by redistributing the very low probability values. More concretely, the nucleus distribution recomputes the probability distribution only on the $k$ most probable terms, where $k$ is chosen such that the accumulated probabilities of these $k$ terms is equal to or greater than a predefined threshold $p$. Similar to \citet{xu2020understanding}, we set $p=0.95$.

Given the nucleus probability distribution for the generation of the term at time step $i$, denoted as $X^{i}$, the term-level uncertainty of the model is calculated as follows:
\begin{equation}
\text{term-level uncertainty}(X^{i}) = -\sum_{x\in X^{i}}P(x) \cdot \log {P(x)}
\end{equation}

Using this definition, we can estimate a model's uncertainty with respect to generating the whole query, namely \textit{query-level uncertainty}, by aggregating term-level uncertainty values. To this end, various aggregation functions (such as mean, entropy, variance, and maximum) can be applied to the corresponding values of each query. We further investigate the characteristic of this uncertainty estimation for model/collection analysis and the cutoff prediction task in Section~\ref{sec:results:uncertainty} and Section~\ref{sec:results:cutoff}, respectively.







\vspace{-0.2cm}
\section{Experiment Setup}
\label{sec:setup}
\paragraph{Collections.} 
We conduct our evaluation experiments on two paragraph retrieval collections. The first is the MS~MARCO Passage Re-ranking collection~\cite{nguyen2016ms}. In total, the development set of MS~MARCO comprises 8,841,822 documents, and 55,578 queries. We follow the setting in \citet{hofstatter2019effect} and split the queries into a validation and a test set, containing 6,980 and 48,598 queries, respectively. Since the provided relevance judgements by this collection are highly sparse, we refer to this test set as SPARSE. The second test collection is the TREC Deep Learning Track 2019 Passage Retrieval set (TREC-19)~\cite{craswell2020overview}, which also originates from the MS~MARCO collection. The TREC-19 collection encompasses 43 annotated queries. 

\begin{table*}[t]
\begin{center}
\caption{\label{tbl:results} 
Results of investigated models in terms of MRR, NDCG, and Recall. 
Best performances among generative models are marked in bold. Superscripts show significant improvement over respective models trained with the same loss.
}
\centering
\scalebox{0.96}{
\begin{tabular}{l L{1.2cm} | l l l | l l l}
\toprule
\multirow{2}{*}{Model}  & \multirow{2}{*}{Loss}  & \multicolumn{3}{c}{\textbf{SPARSE}} & \multicolumn{3}{c}{\textbf{TREC-19}} \\
 &  & MRR & NDCG & Recall & MRR & NDCG & Recall \\\midrule
 
BM25 &   & $0.199$ & $0.231$ & $0.383$ & $0.825$ & $0.506$ & $0.129$ \\
\modelmp  & \lossmxmrg & $0.242$ & $0.280$ & $0.450$ & $0.884$ & $0.577$ & $0.135$ \\
\modelknrm  & \lossmxmrg & $0.234$ & $0.274$ & $0.448$ & $0.861$ & $0.545$ & $0.138$ \\
\modelcknrm  & \lossmxmrg & $0.275$ & $0.318$ & $0.498$ & $0.901$ & $0.605$ & $0.152$ \\
\modeltk  & \lossmxmrg & $0.308$ & $0.355$ & $0.545$ & $0.943$ & $0.661$ & $0.159$ \\
\modelbert  & \lossce & $0.305$ & $0.353$ & $0.542$ & $0.899$ & $0.651$ & $0.152$ \\\midrule
QL ($a$) &   & $0.181$ & $0.211$ & $0.355$ & $0.773$ & $0.470$ & $0.124$ \\\cdashlinelr{1-8}
\multirow{3}{*}{\modelss ($b$)}  & \lossnegl  & $0.246^{a}$ & $0.285^{a}$ & $0.455^{a}$ & $0.825$ & $0.557^{a}$ & $0.141$ \\
  & \lossmxmrg  & $0.210^{a}$ & $0.243^{a}$ & $0.399^{a}$ & $0.860$ & $0.530$ & $0.123$ \\
 &  \losslul  & $0.243^{a}$ & $0.282^{a}$ & $0.453^{a}$ & $0.859$ & $0.558^{a}$ & $0.140$ \\\cdashlinelr{1-8}
\multirow{3}{*}{\modeltt ($c$)}  & \lossnegl  & $0.255^{ab}$ & $0.297^{ab}$ & $0.478^{ab}$ & $0.846$ & $0.541$ & $0.148$ \\
 & \lossmxmrg  & $0.258^{ab}$ & $0.299^{ab}$ & $0.474^{abd}$ & $0.893$ & $0.590^{ab}$ & $0.138$ \\
 & \losslul  & $0.252^{ab}$ & $0.295^{ab}$ & $0.475^{ab}$ & $0.883$ & $0.544$ & $0.142$ \\\cdashlinelr{1-8}
\multirow{3}{*}{\modelbertt ($d$)}   & \lossnegl  & $0.257^{abc}$ & $0.300^{abc}$ & $0.480^{ab}$ & $0.831$ & $0.554^{a}$ & $0.149^{b}$ \\
 & \lossmxmrg  & $0.257^{ab}$ & $0.297^{ab}$ & $0.469^{ab}$ & $0.863$ & $0.573^{a}$ & $0.136$ \\
 & \losslul  & $0.258^{abc}$ & $0.300^{abc}$ & $0.478^{ab}$ & $0.873$ & $0.571^{a}$ & $\textbf{0.150}^{abc}$ \\\cdashlinelr{1-8}
\multirow{3}{*}{\modelpgn ($e$)}   & \lossnegl  & $0.273^{abcd}$ & $0.317^{abcd}$ & $0.498^{abcd}$ & $0.907^{a}$ & $0.585^{acd}$ & $0.150^{b}$ \\
  & \lossmxmrg  & $0.275^{abcd}$ & $0.317^{abcd}$ & $0.493^{abcdf}$ & $\textbf{0.912}^{a}$ & $\textbf{0.609}^{ab}$ & $0.145^{b}$ \\
 &  \losslul  & $0.272^{abcd}$ & $0.316^{abcd}$ & $0.498^{abcd}$ & $0.845$ & $0.569^{a}$ & $0.149^{b}$ \\\cdashlinelr{1-8}
\multirow{3}{*}{\modeltpgn ($f$)}  & \lossnegl  & $0.278^{abcde}$ & $0.323^{abcde}$ & $0.506^{abcde}$ & $0.885$ & $0.575^{a}$ & $0.144$ \\
 & \lossmxmrg  & $0.276^{abcd}$ & $0.317^{abcd}$ & $0.488^{abcd}$ & $0.880$ & $0.601^{ab}$ & $0.148^{ab}$ \\
 & \losslul  & $\textbf{0.281}^{abcde}$ & $\textbf{0.325}^{abcde}$ & $\textbf{0.508}^{abcde}$ & $0.891^{a}$ & $0.573^{a}$ & $0.145$ \\
\bottomrule
\end{tabular}
}
\end{center}
\vspace{-0.3cm}
\end{table*}

\vspace{-0.2cm}
\paragraph{Generative deep ranking models.} Within the proposed generative neural re-ranking framework, we investigate \modelss, \modeltt, \modelbertt, \modelpgn, and \modeltpgn.

\vspace{-0.2cm}
\paragraph{Matching models.}
For the sake of a well rounded performance evaluation, we sample a number of IR models to compare generative models to: Kernel-based Neural Ranking Model (\modelknrm)~\cite{xiong2017end} Convolutional KNRM (\modelcknrm)~\cite{dai2018convolutional}, \modelmp~\cite{pang2016text}, and two most recent Transformer-based models: Transformer-Kernel (\modeltk)~\cite{Hofstaetter2020_sigir} and the fine-tuned \modelbert model~\cite{devlin2019bert}. This list is indeed non-comprehensive, since our central aim is to investigate model architectures within the neural generative paradigm. We conduct experiments on BM25 as a classical matching model.

\vspace{-0.2cm}
\paragraph{Model configuration and training.}  
To provide a fair comparison, we aim to select similar configurations for the different models. Every model using pre-trained word embeddings (\modelss, \modelpgn, \modeltpgn, \modeltt, and \modeltk) operates with the same set of pre-trained GloVe~\cite{pennington2014glove} vectors of length 300. For models with BERT (\modelbert, \modelbertt), we investigate a recently-released version of the pre-trained language model known as BERT-Tiny~\cite{turc2019wellread}, which has two layers of Transformers, two attention heads on each layer, an intermediate feed-forward layer of size 512, and a final (sub)word embedding of size 128. While much smaller, this model has shown competitive performance in various language processing tasks~\cite{turc2019wellread} in comparison with the larger versions of BERT, making it suitable for conducting large-scale experiments with various hyper-parameter settings. We set the setting of all other model with Transformer networks (\modeltk, \modeltpgn, and \modeltt) to the same one as BERT-Tiny. Models that contain BERT (\modelbert, \modelbertt) utilize WordPiece tokenization. All state-of-the-art models are trained with their recommended loss functions, namely Cross Entropy (\lossce) for \modelbert and \lossmxmrg for the other matching models. The proposed generative models are trained using three different loss functions: \lossnegl, \lossmxmrg and \losslul. \modelss and \modelpgn have a learning rate of 0.001. Non-BERT Transformer-based models start from learning rates of 0.0001. BERT-based generative models use a learning rate of 0.0001 for training the Transformer-decoder, and 0.00003 for the pre-trained BERT encoder. The complete hyperparameter settings of all models are provided in the published repository together with the source code.\footnote{\url{https://github.com/CPJKU/DeepGenIR}.}


\vspace{-0.2cm}
\paragraph{Evaluation.} We evaluate the performance of all models based on the re-ranking approach. To this end, we first compute the top 200 passages as retrieved by a BM25 model. The resulting candidate documents are then re-ranked by each of the investigated neural model. The final re-ranked results are evaluated using several common performance metrics, namely mean reciprocal rank (MRR), normalized discounted cumulative gain at 10 (NDCG), and recall. To investigate statistical significance of results, we conduct two-sided paired $t$-tests (details given below). In addition, we qualitatively analyze a selection of generated queries.

\vspace{-0.2cm}
\section{Results and Analysis}
\label{sec:results}
In this section, we first show the performance evaluation results of the various deep generative models, followed by qualitative analysis of the query generation process. We then explore the use of uncertainty estimates to analyze the underlying characteristics of the model and data, followed by showing the benefit of including uncertainty information in the cut-off prediction task.

\vspace{-0.2cm}
\subsection{Performance Evaluation}
Evaluation results are provided in Table~\ref{tbl:results} for all assessed models. Matching models are grouped at the top of the table, and the lower part is dedicated to generative models. For each neural generative model, the results on three loss functions (\lossnegl, \lossmxmrg, \losslul) are reported. The best performance among all generative models is marked in bold. To denote statistical significance, we first assign each generative model a letter $a$ to~$f$ (see first column of Table~\ref{tbl:results}). Each performance result of each model is also marked with superscript letters, indicating to which other models a statistically significant difference exists. To give an example: model \modeltpgn trained with loss \lossnegl, obtaining a MRR of $0.278^{abcde}$ on the SPARSE test set, is significantly better (in terms of MRR) than generative models $a$, $b$, $c$, $d$ and $e$ which have also been trained with the same loss \lossnegl.

Let us have a closer look at the results of generative models. The results indicate that the models that use the copy mechanism show the best overall performance among the generative models. In particular, \modeltpgn shows significantly better results than all other deep generative models on SPARSE, while \modelpgn shows better performance on TREC-19. The better performance of PGN-based models (\modelpgn and \modeltpgn) in contrast to BERT-based ones is specific to the retrieval task, and in fact stands in contrast to the common architectural preferences in other tasks such as machine translation and abstractive document summarization. 

The effectiveness of the PGN-based models can be traced in their decoder architectures, particularly by comparing between \modelpgn and \modelss. While the sole difference of these two models lies in the use of the copy mechanism, the \modelpgn and \modeltpgn models show significantly higher results with large margins. We assume that this is due to the way that the copy mechanism in PGN-based models approach out-of-vocabulary terms (OOVs). In fact, as observed in previous studies~\cite{hofstatter2019effect}, OOVs correspond to infrequent words that -- due to their rarity -- contain crucial information for retrieval. Models that leverage this information, therefore, reach higher performance levels. While \modelpgn and \modeltpgn both benefit from effective decoding (in respect to these retrieval tasks), the improvement of \modeltpgn on SPARSE highlights the importance of enriching the encoding layer with Transformers which differentiates the \modeltpgn model from \modelpgn. 


Inspecting results for the different loss functions used for the deep generative models reveals that, overall, the differences between various loss functions are negligible, such that the models using \lossnegl (as the simplest loss function) perform generally similar to the ones with \lossmxmrg or \losslul. We speculate that this is due to the probabilistic nature of generative models, as the objective of such models is to estimate generation probability distributions, which (based on the results) can be achieved by solely increasing the generation probability of relevant documents. We therefore conclude that a generative model can effectively be trained with the \lossnegl loss function as the simplest choice, which has the benefit of faster training time in comparison with other loss functions.\footnote{Since \lossnegl in contrast to \lossmxmrg and \losslul only processes the relevant documents.} Finally, comparing the results of deep generative models with the state-of-the-art query-document interaction models with Transformers and BERT, we observe that overall the generative models show only marginally lower performance.\footnote{Comparing the latency of models, it is expected that the neural generative models have overall longer inference time due to their generation process. In particular, we observe that the PGN-based models, due to the use of two LSTMs at encoder and decoder, have considerably longer inference time. However, \modelbertt, while performing marginally lower than the PGN-based models, shows almost on-par latency to the \modelbert ranker.}

These observations on the significant differences between various architectural choices are particularly important considering that, as discussed in Section~\ref{sec:related}, most current studies which exploit generative models (e.g.,~for tasks such as query reformulation) use similar models to \modelss~\cite{zamani2020generating,yang2019hybrid,ren2018conversational} or the ones that utilize Transformers as decoder~\cite{nogueira2019document}. Based on our results, exploiting OOV-aware models such as \modeltpgn can provide considerable benefits for the corresponding final tasks.
\vspace{-0.2cm}




\begin{table}[t]
\begin{center}
\caption{\label{tbl:examples} 
Examples of passages, actual queries for which the passage was marked relevant, and synthetic queries most likely to be generated by \modeltpgn. 
}
\vspace{-0.2cm}
\scriptsize
\begin{tabular}{p{8cm}}
\toprule
\multicolumn{1}{c}{\textbf{Example 1}}\\
\textbf{Passage}: Fleas are holometabolous insects, going through the four lifecycle stages of egg, larva, pupa, and imago (adult). Adult fleas must feed on blood before they can become capable of reproduction. Flea populations are distributed with about 50\% eggs, 35\% larvae, 10\% pupae, and 5\% adults. Generally speaking, an adult flea only lives for 2 or 3 months. Without a host for food a flea's life might be as short as a few days. With ample food supply, the adult flea will often live up to 100 days.\\
\textbf{Actual query}: how long is life cycle of flea \\ 
\textbf{Generated query}: how long do fleas live \\ 
\midrule
\multicolumn{1}{c}{\textbf{Example 2}}\\ 
\textbf{Passage}: I have always wanted to become a Nurse and I have been doing some research and came across the different Nursing 'titles' such as RN (Registered Nurse), BSN(Bachlor's in Science of Nursing) NA(Nurse Assistant), CRNA (Certified Registered Nurse Anesthetist), LPN and LVN.SN = Bachelor of Science in Nursing, which is just a 4 year RN degree. Both the 2 year and the BSN graduates sit for the exact same licensure exam and earn the same RN license.\\
\textbf{Actual query}: difference between rn and bsn \\ 
\textbf{Generated query}: what degree do you need to be a nurse \\
\midrule
\multicolumn{1}{c}{\textbf{Example 3}}\\ 
\textbf{Passage}: The flea population is typically. made up of 50\% eggs, 30\% larvae, 15\% pupae and only 5\% biting adults. Completion of the life cycle from egg to adult varies from two weeks to eight months. Normally the female flea lays about 15 to 20 eggs per day up to 600 in a lifetime. Usual hosts for fleas are dogs, cats, rats, rabbits, mice, squirrels, chipmunks, raccoon's, opossums, foxes, chickens, and humans. \\
\textbf{Actual query}: how long is life cycle of flea \\ 
\textbf{Generated query}: how long do chickens live \\
\bottomrule
\end{tabular}
\label{tbl:results:examples} 
\end{center}
\vspace{-0.2cm}
\end{table}

\subsection{Qualitative analysis of generated queries.}
We now look at the query generation aspect of the models from a qualitative perspective. In the current and next section, we use \modeltpgn as our overall best-performing deep generative model to generate queries in a greedy generation process. In this process, for every position the token with the highest probability is selected from the generation probability distribution of words. The generated query in this way is a greedy approximation of the query with the highest generation probability for the given passage. 


Table~\ref{tbl:examples} shows examples of passages, provided queries in the dataset assessed as relevant to the corresponding passages, and the queries generated by our model. We expect that a generated query is conceptually relevant to the given passage. Looking at Example~1, we observe that the generated query is almost the same as the actual relevant one. It means that the model will predict a high relevance score of the query to the passage. Example~2 shows the opposite situation: the generated query, while being completely different from the actual one, is still a valid and relevant query to the given passage. Finally, in Example~3, the same actual query as the one in Example~1 is used but with a different (while still relevant) passage. This example highlights a failure case where the generated query (according to the discussed greedy approach) is conceptually non-relevant to the passage. These examples motivate the deeper understanding of neural generative ranking models, while the shown cases are directly relevant to the tasks that exploit query generation for downstream tasks~\cite{nogueira2019document}.

\begin{figure}[t]
\centering
\includegraphics[width=0.37\textwidth]{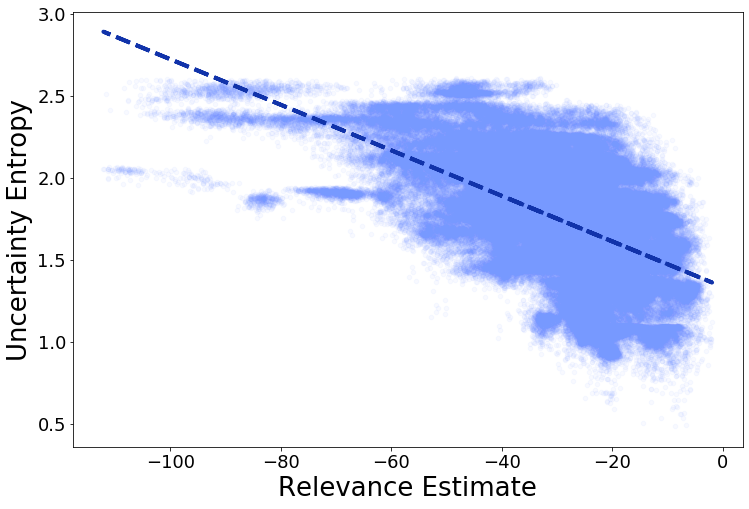}
\centering
\caption{\label{fig:correlation}Relevance versus query-level uncertainty of the T-PGN model on TREC-19.}
\vspace{-0.4cm}
\end{figure}

\vspace{-0.2cm}
\subsection{Model Understanding Through the Lens of Uncertainty}
\label{sec:results:uncertainty}
In the following, we present model- and data-related insights we obtained from analyzing uncertainty estimates for the T-PGN model (Section~\ref{sec:method_uncertainty}) to approach the following questions: (1) Is there any connection between the model's confidence in its query generation probability and query-document relevance estimates? (2) Are there any patterns in the uncertainty distribution along query term positions and, if yes, what do they indicate? 

To address the first question, we start with calculating query-level uncertainty estimates by aggregating over term-level uncertainties using mean, entropy, variance, and maximum. Then, for each query-document pair in the top-200 of a ranking list, we calculate the Spearman-$r$ correlation between each query-level uncertainty and the predicted relevance scores. We calculate these correlations for TREC-19, containing $8,\!600$ query-document pairs. 

The calculated correlations for mean, variance, max, and entropy are -0.223, -0.206, -0.358, and -0.569, respectively. All different uncertainty aggregation results show a negative correlation to relevance score, indicating that a decreasing predicted relevance score (for the documents in the lower positions in the ranking list) increases uncertainty of the model. 
Figure~\ref{fig:correlation} demonstrates the relevance and query-level uncertainty estimates using entropy for aggregation, because of its highest negative correlation. The plot shows that uncertainty of query-document pairs with higher relevance is widely spread, but the distribution tends to get focused on a high-uncertainty area as the relevance decreases. 

Considering these results, with regard to our first question, we conclude that the model tends to exhibit higher levels of uncertainty (in the likelihood of generating the query in query-document pairs) for low relevance estimates. This could indicate that uncertainty may contain additional information to relevance which can be exploited in retrieval tasks. We return to this point in Section~\ref{sec:results:cutoff}.

\begin{figure}[t]
\centering
\includegraphics[width=0.39\textwidth]{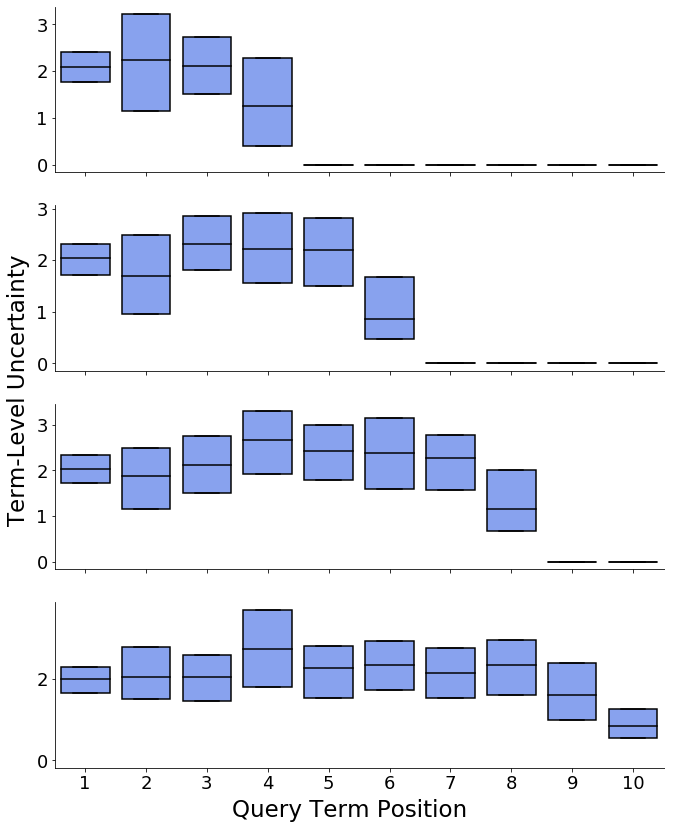}
\centering
\caption{\label{fig:UncertaintyDist}The interquartile ranges of term-level uncertainty scores, calculated on each term position for all queries with a given length, namely the length of 4, 6, 8, and 10. For each query length, the last term position corresponds to the $\texttt{<EOS>}$ special token denoting the end of the query.}
\end{figure}

To approach the second question, using the query-document pairs of the TREC-19 results, we average the term-level uncertainty values over all query terms that appear in a specific position of the queries. To make the results comparable across various query lengths, we apply this position-level aggregation over the queries with the same size. Figure~\ref{fig:UncertaintyDist} shows these term-level uncertainty distributions for every position in queries of length 4, 6, 8, and 10 terms, where each query ends with the $\texttt{<EOS>}$ special token. Every box in the plot represents the interquartile range of term-level uncertainty distribution for each position. 

Looking at Figure~\ref{fig:UncertaintyDist}, we observe two major patterns in all four settings regarding query length: (1) All average uncertainties tend to become lower (more confident) in the last terms of the query, where the last term has consistently the lowest uncertainty with a considerable drop in comparison to the uncertainty at previous term positions. (2) The uncertainty distributions regarding the first position have similar median values across the four settings with small variances when compared to the distributions for other positions.

Our first observation is similar to the findings by \citet{xu2020understanding} in the context of abstractive text summarization. This indicates that by observing more terms during the query generation, the model becomes more and more certain about the distribution of possible next terms, and this confidence has its maximum in the last term.

Our second observation is, however, in contrast to the results reported by \citet{xu2020understanding}. 
In their experiments, generative models show the greatest interquartile range of term-level uncertainty for earlier words in the generated sequence. This can potentially reveal the existence of a bias in the queries of the MS~MARCO training dataset, considering that many queries in the dataset start with question words such as \emph{what}, \emph{how}, and \emph{where}. In fact, the persistent uncertainty distributions for the first position can indicate the limited number of unique terms in training data, with which a query begins. This observation is inline with and reinforces the conclusions in \citet{hofstatter2020fine}. However, while they show the existence of bias in the MS~MARCO collection through extensive fine-grained annotation, we view this from the lens of the uncertainty of the model on each query term position.



\vspace{-0.2cm}
\subsection{Cut-off Prediction with Uncertainty}
\label{sec:results:cutoff}
Do the uncertainty estimates provide novel and complementary information to what is provided by relevance scores? If yes, can this information be exploited in downstream IR tasks?
To answer these questions we evaluate the expressiveness of the uncertainty estimates in a similar fashion to Cohen et al.~\citep{cohen21notall}, via the cutoff prediction task.
The objective of the cut-off prediction task is to dynamically determine a cut-off point for a ranked list of documents in order to maximize some non-monotonic metric, in our case $F_1$ scores. As discussed by \citet{bicut_cutoffLien19}, the task is motivated by neural models losing confidence in their estimations as documents become less relevant to the query. In a real-world scenario, cut-off prediction can be used by a retrieval system to prevent users from scraping over search results, about which the ranker is not sufficiently confident. In such scenarios, the search engine can switch to alternative strategies, such as applying different ranking model or encouraging the user to reformulate the query. 

To study the effect of uncertainty on this task, we follow the same procedure as in \citet{choppycutoffBahri20}, namely by using the proposed Transformer-based cut-off predictor, and comparing the performance in terms of $F_1$ score (see \citet{choppycutoffBahri20} for more details). The predictor receives a set of features in the sense of query-document interactions, and for each query provides a prediction regarding the best cut-off in its ranked list. A common feature for this task is relevance scores, assuming 
that the changes in relevance can be indicative of an optimal cut-off point~\cite{choppycutoffBahri20}. In our experiments, we are interested in examining whether adding uncertainty information can further improve this tasks by providing new information.

We therefore conduct our experiments in two configurations: (1) using only relevance estimation from T-PGN as single feature, referred to as \texttt{Rel}; (2) adding the four query-level uncertainty estimates (through mean, entropy, variance, and maximum term-level uncertainty aggregations) as additional features, referred to as \texttt{Rel+Uncertainty}. To train the cut-off predictors 
we use the queries of TREC-19. While this task can benefit from the large number of the queries in SPARSE, the task intrinsically requires a sufficient amount of relevance judgements which are not available in the SPARSE collection. In addition to \texttt{Rel} and \texttt{Rel+Uncertainty}, we calculate the results of a \texttt{Greedy} approach which provides a naive baseline by selecting the same cut-off for all ranked lists, chosen by maximizing the $F_1$ score on the training set. 
Finally, the \texttt{Oracle} model indicates the score that an ideal cut-off selection would achieve. We report in Table~\ref{tbl:cutoff}, for each configuration, the resulting $F_1$ score as well as its percentage when compared to the $F_1$ score 
of \texttt{Oracle}. For each of the configuration, the experiment is conducted in 50 trials, where in each trial 5-fold cross validation is applied. The final results are averaged over all trials.

Comparing results for \texttt{Rel} and \texttt{Greedy} in Table~\ref{tbl:cutoff} -- as reported in previous studies~\citet{choppycutoffBahri20,bicut_cutoffLien19} -- we observe that relevance information is an important signal for this task. Comparing the results of \texttt{Rel} with \texttt{Rel+Uncertainty} we observe additional improvements by incorporating the uncertainty information. Calculating a two-sided t-test with $p<0.001$ between the results of \texttt{Rel+Uncertainty} and \texttt{Rel} confirms the significance of this improvement. These results substantiate the value of the uncertainty scores, inherent in the architecture of deep generative IR models, which provide additional actionable information for IR tasks. 

\begin{table}
\caption{ \label{tbl:cutoff}
Results on cut-off prediction task with features produced by T-PGN on TREC-19 test collection. The last column show the percentage of $F_1$ in respect to the results of \texttt{Oracle}. The $\dagger$ sign shows the statistical improvement of \texttt{Rel+Uncertainty} over \texttt{Rel} with $p<0.001$.
}
\vspace{-0.2cm}
\begin{tabular}{llc}
\toprule
& \multicolumn{1}{c}{\textbf{$F_1$}} & \textbf{\% to \texttt{Oracle}} \\
 \hline
\texttt{Greedy} & 0.193 & 39.1 \\
\texttt{Oracle} & 0.493 & 100.0 \\\midrule
\texttt{Rel} & 0.345 & 70.0 \\
\texttt{Rel+Uncertainty} & $0.364\dagger$ & 73.8 \\
 \bottomrule
\vspace{-0.6cm}
\end{tabular}
\end{table}

\vspace{-0.2cm}
\section{Conclusion}\label{sec:conclusions}
We propose a modern perspective on the generative IR paradigm by introducing novel deep generative ranking models. The introduced models offer a solid granular probabilistic framework of neural retrieval, which lays the foundation for estimation of additional model-level information such as uncertainty. Proposing a novel deep generative ranking model, \modeltpgn, we investigate the performance of several deep generative IR models on two passage retrieval collections. Our evaluation results show the importance of the copy mechanism in the generative models in the context of retrieval, as provided by the \modelpgn and \modeltpgn models. We further explore the information provided by the uncertainty estimates, and showcase the value of such uncertainty information in a cut-off prediction task.

\vspace{-0.2cm}
\section*{Acknowledgements}
This research is supported in part by the NSF (IIS-1956221). The views and conclusions contained herein are those of the authors and should not be interpreted as necessarily representing the official policies, either expressed or implied, of NSF or the U.S. Government. This research is also supported by Know-Center Graz, through project “Theory-inspired Recommender Systems”.



\bibliographystyle{ACM-Reference-Format}
\bibliography{refs}

\end{document}